\documentclass[prl,twocolumn,showpacs,amsmath,amssymb]{revtex4}

\usepackage{graphicx}
\usepackage{dcolumn}
\usepackage{bm}

\begin{document}
\title
{Theory of Flux-Flow Resistivity near $H_{c2}$ for $s$-wave Type-II Superconductors}

\author{Takafumi Kita}
\affiliation{Division of Physics, Hokkaido University, Sapporo 060-0810,
Japan}

\date{\today}

\begin{abstract}
This paper presents a microscopic calculation of the
flux-flow resistivity $\rho_{f}$ for $s$-wave type-II superconductors
with arbitrary impurity concentrations near the upper critical field $H_{c2}$.
It is found that, as the mean free path $l$ becomes longer, $\rho_{f}$
increases gradually from the dirty-limit result of 
Thompson [Phys.\ Rev.\ B{\bf 1},\ 327 (1970)] and 
Takayama and Ebisawa [Prog.\ Theor.\ Phys.\ {\bf 44},\ 1450 (1970)].
The limiting behaviors suggest that $\rho_{f}(H)$ at low temperatures
may change from convex downward to upward as $l$ increases,
thus deviating substantially from the linear dependence $\rho_{f}\!\propto\! H/H_{c2}$ 
predicted by the Bardeen-Stephen theory [Phys.\ Rev.\ {\bf 140},\ A1197 (1965)].
\end{abstract}
\maketitle

Kim {\em et al}.\ \cite{KS69} attributed finite resistivity
observed in type-II superconductors
to the motion of flux lines, calling it ``flux-flow resistivity'' $\rho_{f}$.
Whereas the idea has been accepted widely,
we still have poor quantitative understanding of the phenomenon.
The early phenomenological theories 
based on single-flux considerations \cite{Tinkham64,BS65,NV66} cannot
explain the steep decrease of $\rho_{f}$ 
observed near $H_{c2}$ \cite{KS69}.
This region,
where microscopic calculations may be performed most easily, 
has been a subject of later theoretical works
\cite{Schmid66,CM67,Thompson70,TE70}.
We thereby have a quantitative theory
on the flux-flow resistivity near $H_{c2}$ \cite{Thompson70,TE70},
but its validity is still restricted to the dirty limit.
Reasonable agreements with the theory have been reported later in a couple of 
experiments on microwave surface resistivity \cite{PKT73}
and flux-flow resistivity \cite{Gey76}.
However, the latter experiment found a small discrepancy that $\rho_{f}$ near $T_{c}$
is larger than the theoretical prediction \cite{Thompson70,TE70},
which was attributed by Larkin and Ovchinnikov \cite{Larkin86} to the extra
phonon scattering effective at finite temperatures.

With these backgrounds, this paper provides a quantitative theory 
on the flux-flow resistivity near $H_{c2}$ applicable to arbitrary impurity concentrations.
I thereby hope to establish the domain in which the dirty-limit theory 
\cite{Thompson70,TE70} is valid,
and look into whether the experiments \cite{PKT73,Gey76} 
may be explained by impurity scattering alone.
Although a similar study was carried out by Ovchinnikov \cite{Ovchinnikov74},
the behaviors of $\rho_{f}$
at intermediate impurity concentrations have not been clarified explicitly.

I consider the $s$-wave pairing with an isotropic Fermi surface and the $s$-wave 
impurity scattering in an external magnetic field ${\bf H}\!\parallel \!{\bf z}$.
I calculate the complex conductivity $\sigma(\omega)$ 
below microwave frequencies $\omega$
using the quasiclassical equations of superconductivity \cite{Rainer83},
and finally put $\omega\!\rightarrow\!0$.
I use the notation of Ref.\ \onlinecite{Kita01} but
leave the Hall terms out of consideration.
Hence I start from the same equations 
as Eschrig {\em et al}.\ \cite{Eschrig99} in clarifying the motion of
a single flux line within the linear-response regime.

The vector potential without perturbation is given by
${\bf A}({\bf r})\!=\! Bx\hat{\bf y}\!+\tilde{\bf A}({\bf r})$,
where $B$ is the average flux density and
$\tilde{\bf A}$ expresses the spatially varying part of the magnetic field 
satisfying $\int{\bm \nabla}\!\times\!\tilde{\bf A}\,d{\bf r}\!=\!{\bf 0}$.
The corresponding retarded quasiclassical Green's functions $f^{\rm R}$ and $g^{\rm R}$
can be obtained as Ref.\ \onlinecite{Kita03} with the
replacement of the Matsubara frequency $\varepsilon_{n}$ by $-i\varepsilon$.
I adopt the units where the energy, length, magnetic field, and electric field
are measured by the zero-temperature energy gap $\Delta(0)$ at $H\!=\!0$,
the coherence length
$\xi_{0}\!\equiv\!\hbar v_{\rm F}/\Delta(0)$ with $v_{\rm F}$ the Fermi velocity, 
$B_{0}\!\equiv\!\phi_{0}/2\pi\xi_{0}^{2}$ with $\phi_{0}\!\equiv\! hc/2e$ the flux quantum, 
and $E_{0}\!\equiv\!\hbar v_{\rm F}/e\xi_{0}^{2}$,
respectively. I also put $\hbar\!=\! k_{\rm B}\!=\! 1$.

Now, consider the response to a spatially uniform but time-dependent
perturbation $\delta\hspace{-0.3mm}{\bf A}{\rm e}^{-i\omega t}\!=\!
\delta\hspace{-0.3mm}{\bf E}\,{\rm e}^{-i\omega t}/i\omega$
with $\delta\hspace{-0.3mm}{\bf E}\!\perp\!{\bf H}$.
A straightforward calculation based on Eq.\ 
(71) of Ref.\ \onlinecite{Kita01}
leads to the following equation 
for the first-order response 
$\delta\hspace{-0.3mm} f^{\rm R}\!=\!
\delta\hspace{-0.3mm}f^{\rm R}(\varepsilon,{\bf k},\omega,{\bf r})$:
\begin{subequations}
\label{Eilen-R}
\begin{eqnarray}
&& \left[-2i\varepsilon+\frac{\langle g^{\rm R}_{+}\rangle+\langle g^{\rm R}_{-}\rangle}{2\tau}
+\hat{\bf v}\!\cdot\!\left({\bm \nabla}-i{\bf A}\right)\right] 
\delta\hspace{-0.3mm} f^{\rm R}
\nonumber\\
&& \hspace{-3mm}
=( f^{\rm R}_{+}+f^{\rm R}_{-})\,{\hat{\bf v}\!\cdot\!\delta\hspace{-0.3mm}{\bf E}}/{\omega}\,
+( g^{\rm R}_{+}+g^{\rm R}_{-})\,\delta\hspace{-0.3mm}\Delta
\nonumber\\
&&+
\left(\Delta+\frac{\langle f^{\rm R}_{+}\rangle}{2\tau}\right)
\delta g^{\rm R\dagger}
 +\left(\Delta+\frac{\langle f^{\rm R}_{-}\rangle}{2\tau}\right)
\delta g^{\rm R}
\nonumber\\
&& -\frac{f^{\rm R}_{+}\langle \delta g^{\rm R\dagger}\rangle
+f^{\rm R}_{-}\langle \delta g^{\rm R}\rangle}{2\tau}
+\frac{( g^{\rm R}_{+}+g^{\rm R}_{-})\langle\delta\hspace{-0.3mm} f^{\rm R}\rangle}{2\tau}
\, .
\label{Eilen-fR}
\end{eqnarray}
Here $\tau$ is the relaxation time in the Born approximation,
$\langle \cdots \rangle$ denotes
the Fermi-surface average satisfying $\langle 1 \rangle\!=\! 1$, 
and $\Delta({\bf r})$ and $\delta\hspace{-0.3mm}\Delta(\omega,{\bf r})$
are the pair potential and its first-order response, respectively.
The unit vector $\hat{\bf v}\!=\!\hat{\bf k}$ 
specifies a point on the spherical Fermi surface,
$f^{\rm R}_{\pm}\!\equiv\! f^{\rm R}(\varepsilon_{\pm},{\bf k},{\bf r})$
and $g^{\rm R}_{\pm}\!\equiv\! g^{\rm R}(\varepsilon_{\pm},{\bf k},{\bf r})$
with $\varepsilon_{\pm}\!\equiv\!\varepsilon\!\pm\!{\omega}/{2}$,
and the superscript $^{\dagger}$ denotes simultaneous operations
of complex conjugation
and $(\varepsilon,{\bf k},\omega)\!\rightarrow\!(-\varepsilon,-{\bf k},-\omega)$,
e.g., $\delta g^{\rm R\dagger}(\varepsilon,{\bf k},\omega,{\bf r})\!=\!
\delta g^{\rm R*}(-\varepsilon,-{\bf k},-\omega,{\bf r})$.
Finally, the normalization condition \cite{Rainer83} enables us to write 
$\delta g^{\rm R}$ in terms of $\delta\hspace{-0.3mm}f^{\rm R}$ as
\begin{equation}
\delta g^{\rm R}=-\bigl(\,f^{\rm R}_{+}\,\delta\hspace{-0.3mm} f^{\rm R\dagger}
+ f^{\rm R\dagger}_{-}\,\delta\hspace{-0.3mm} f^{\rm R}
\,\bigr)/
\bigl(\,g^{\rm R}_{+}+ g^{\rm R}_{-}\,\bigr) \, .
\label{Eilen-gR}
\end{equation}
\end{subequations}
Equation (\ref{Eilen-R}) determines the retarded functions $\delta\hspace{-0.3mm} f^{\rm R}$
and $\delta g^{\rm R}$ for given $\delta\hspace{-0.3mm}{\bf E}$ and 
$\delta\hspace{-0.3mm}\Delta$. 
In addition,
the advanced functions are obtained directly by using Eq.\ (72) of Ref.\ \onlinecite{Kita01} as
$\delta\hspace{-0.3mm} f^{\rm A}(\varepsilon,{\bf k},\omega,{\bf r})\!=\!
\delta\hspace{-0.3mm} f^{\rm R}(-\varepsilon,-{\bf k},\omega,{\bf r})$ and 
$\delta g^{\rm A}(\varepsilon,{\bf k},\omega,{\bf r})\!=\!
-\delta\hspace{-0.3mm} g^{\rm R*}(\varepsilon,{\bf k},-\omega,{\bf r})$.

As for the Keldysh functions, I write them following Eschrig {\em et al}.\ \cite{Eschrig99} as 
$\delta g^{\rm K}\!=\!\delta g^{\rm R}\phi_{-}\!-\!
\phi_{+}\delta g^{\rm A}\!+\!\delta g^{\rm a}$ and 
$\delta\hspace{-0.3mm} f^{\rm K}\!=\!\delta\hspace{-0.3mm}f^{\rm R}\phi_{-}\!-\!
\phi_{+}\delta\hspace{-0.3mm}f^{\rm A}\!+\!\delta\hspace{-0.3mm}f^{\rm a}$
with $\phi_{\pm}\!\equiv\!\tanh(\varepsilon_{\pm}/2T)$.
Then a simplified equation results for $\delta g^{\rm a}\!=\!
\delta g^{\rm a}(\varepsilon,{\bf k},\omega,{\bf r})$ as
\begin{subequations}
\label{Eilen-a}
\begin{eqnarray}
&& \left(-i\omega+\frac{\langle g^{\rm R}_{+}\rangle-\langle g^{\rm A}_{-}\rangle}{2\tau}
+\hat{\bf v}\!\cdot\!{\bm \nabla}\right)
\delta\hspace{-0.3mm} g^{\rm a}
\nonumber\\
&& \hspace{-3mm}
=(\phi_{+}-\phi_{-})\!
\left[-( g^{\rm R}_{+}-g^{\rm A}_{-})\,
{\hat{\bf v}\!\cdot\!\delta\hspace{-0.3mm}{\bf E}}/{\omega}\,
+f^{\rm R}_{+} \, \delta\hspace{-0.3mm}\Delta^{\!\dagger}
-f^{\rm A\dagger}_{-}\, \delta\hspace{-0.3mm}\Delta\, \right]
\nonumber \\
&&+\left(\Delta+\frac{\langle f^{\rm R}_{+}\rangle}{2\tau}\right)
\delta f^{\rm a\dagger}
+\left(\Delta^{*}+\frac{\langle f^{\rm A\dagger}_{-}\rangle}{2\tau}\right)
\delta f^{\rm a}
\nonumber\\
&& -\frac{f^{\rm R}_{+}\langle \delta f^{\rm a\dagger}\rangle+
f^{\rm A\dagger}_{-}\langle \delta f^{\rm a}\rangle}{2\tau}
+\frac{( g^{\rm R}_{+}-g^{\rm A}_{-})\langle\delta\hspace{-0.3mm} g^{\rm a}\rangle}{2\tau}
\, .
\label{Eilen-ga}
\end{eqnarray}
It also follows from the normalization condition that $\delta\hspace{-0.3mm} f^{\rm a}\!=\!
\delta \hspace{-0.3mm} f^{\rm a}(\varepsilon,{\bf k},\omega,{\bf r})$ is 
given in terms of $\delta g^{\rm a}$ by 
\begin{equation}
\delta f^{\rm a}=-\bigl(\,
f^{\rm R}_{+}\,\delta\hspace{-0.3mm} g^{\rm a\dagger}
+f^{\rm A}_{-}\,\delta\hspace{-0.3mm} g^{\rm a}
\,\bigr)/
\bigl(\,g^{\rm R}_{+}-g^{\rm A}_{-}\,\bigr) \, .
\label{Eilen-fa}
\end{equation}
\end{subequations}

Finally, the pair potential $\delta\hspace{-0.3mm}\Delta$ is determined by
\begin{equation}
\delta\hspace{-0.3mm}\Delta=\frac{N(0)V_{0}}{4i}
\int_{-\varepsilon_{\rm c}}^{\varepsilon_{\rm c}}
\hspace{-2mm}d\varepsilon \,\langle\delta f^{\rm R}\phi_{-}\!-\!
\delta f^{\rm A}\phi_{+}\!+\!\delta f^{\rm a}\rangle
\, ,
\label{pair}
\end{equation}
where $N(0)$ is the density of states per spin at the Fermi level,
$V_{0}\!>\!0$ is the $s$-wave pairing interaction, 
and $\varepsilon_{\rm c}$ is the cut-off energy.
Once the solution to Eqs.\ (\ref{Eilen-R})-(\ref{pair}) is obtained self-consistently, 
the transport current  $\delta{\bf j}\!=\!\delta{\bf j}(\omega,{\bf r})$ is calculated by
\begin{equation}
\delta{\bf j}=-\frac{eN(0)v_{\rm F}}{2}
\int_{-\infty}^{\infty}
\hspace{-2mm}d\varepsilon \,\langle\hat{\bf v}\,(\delta g^{\rm R}\phi_{-}\!-\!
\delta g^{\rm A}\phi_{+}\!+\!\delta g^{\rm a})
\rangle
\, .
\label{current}
\end{equation}
The corrections caused by $\delta{\bf j}$ 
to the magnetic field and the charge density 
may be neglected safely for the relevant weak-coupling
superconductors.
Equations (\ref{Eilen-R})-(\ref{current}) are still exact within the linear-response regime
and form a basic starting point to calculate complex conductivity of type-II superconductors
at arbitrary magnetic fields.

I now concentrate on the region near $H_{c2}$ and solve Eqs.\
(\ref{Eilen-R})-(\ref{pair}) by expanding every quantity up to second order
in $\Delta({\bf r})$
as $f^{\rm R}\!=\!f^{{\rm R}(1)}$, 
$g^{\rm R}\!=\! 1\!-\! \frac{1}{2}f^{{\rm R}(1)}f^{{\rm R}(1)\dagger}$, 
$\tilde{\bf A}\!=\!\tilde{\bf A}^{(2)}$,
$\delta\hspace{-0.3mm}\Delta\!=\!\delta\hspace{-0.3mm}\Delta^{(1)}$,
$\delta\hspace{-0.3mm}f^{\rm R,a}\!=\!\delta\hspace{-0.3mm}f^{{\rm R,a}(1)}$, and
$\delta g^{\rm R,a}\!=\!\delta g^{{\rm R,a}(0)}\!+\!\delta g^{{\rm R,a}(2)}$,
with $g^{{\rm R}(0)}\!=\!0$ as seen from Eq.\ (\ref{Eilen-gR}).
The superscript $^{(1)}$
in $\delta\hspace{-0.3mm}f$ and $\delta\hspace{-0.3mm}\Delta$ will be dropped
as it causes no confusions.

The zeroth-order quantity
$\delta g^{{\rm a}(0)}$ is obtained easily from Eq.\ (\ref{Eilen-ga})
as
\begin{equation}
\delta g^{{\rm a}(0)}=-\frac{2\tau}{1-i\omega\tau}
\frac{\phi_{+}-\phi_{-}}{\omega}\,
\hat{\bf v}\!\cdot\!\delta{\bf E} \, .
\label{g^a(0)}
\end{equation}
When put into Eq.\ (\ref{current}), this expression yields 
the normal-state Drude conductivity $\sigma_{n}$,
as it should.

Next, Eq.\ (\ref{Eilen-fR}) is simplified for the first-order 
$\delta\hspace{-0.3mm} f^{\rm R}$ into
\begin{eqnarray}
\left[-i\varepsilon+\frac{1}{2\tau}
+\frac{\sqrt{B}\sin\theta}{2\sqrt{2}}({\rm e}^{-i\varphi} a
-{\rm e}^{i\varphi}a^{\dagger})\right] 
\delta\hspace{-0.3mm} f^{\rm R}
\nonumber \\
=\frac{f^{\rm R}_{+}+f^{\rm R}_{-}}{2}\,
\frac{\hat{\bf v}\!\cdot\!\delta\hspace{-0.3mm}{\bf E}}
{\omega}
+\delta\hspace{-0.3mm}\Delta
+\frac{\langle\delta\hspace{-0.3mm} f^{\rm R}\rangle}{2\tau}
\, ,
\label{Eilen-fR2}
\end{eqnarray}
where $a^{\dagger}$ and $a$ are creation and annihilation operators satisfying
$[a,a^{\dagger}]\!=\!1$ \cite{Kita98,Kita03}, and
$(\theta,\varphi)$ are the polar angles of $\hat{\bf v}$.
Equation (\ref{Eilen-fR2}) can be solved with the Landau-level expansion (LLX) method \cite{Kita98}
by expanding $\delta\hspace{-0.3mm}\Delta$ and $\delta\hspace{-0.3mm}f^{\rm R}$ 
in periodic basis functions of the vortex lattice as
\begin{subequations}
\begin{equation}
\delta\hspace{-0.3mm}\Delta(\omega,{\bf r})
=\sqrt{V}
\sum_{N=0}^{\infty}\delta\hspace{-0.3mm}\Delta_{N}(\omega)\,\psi_{N{\bf q}}({\bf r}) \, ,
\label{pairExpand}
\end{equation}
\begin{equation}
\delta\hspace{-0.3mm}f^{\rm R}(\varepsilon,{\bf k},\omega,{\bf r})
=\sqrt{V}\!\!\!\sum_{m=-\infty}^{\infty}\!
\sum_{N=0}^{\infty}\!\delta\hspace{-0.3mm}f_{mN}^{\rm R}(\varepsilon,\theta,\omega)
\,{\rm e}^{i m\varphi}\,\psi_{N{\bf q}}({\bf r}) 
\, ,
\label{fR-Expand}
\end{equation}
\end{subequations}
where $N$ denotes the Landau level, ${\bf q}$ is an arbitrary chosen magnetic Bloch vector,
$V$ is the volume of the system, and $\psi_{N{\bf q}}({\bf r})$ satisfies
$a\hspace{0.2mm}\psi_{N{\bf q}}\!=\!\sqrt{N}\psi_{N-1{\bf q}}$ and
$a^{\dagger}\psi_{N{\bf q}}\!=\!\sqrt{N\!+\! 1}\psi_{N+1{\bf q}}$.
The quantities without perturbations are expanded similarly.
Near $H_{c2}$, $\Delta({\bf r})$ may well be approximated using only the $N\!=\! 0$ level
as $\Delta({\bf r})\!=\!\sqrt{V}\Delta_{0}\,\psi_{0{\bf q}}({\bf r})$.
The coefficients $\Delta_{0}$ and $f_{mN}^{{\rm R}}$ 
have already been obtained in Ref.\ \onlinecite{Kita03},
satisfying $f_{mN}^{{\rm R}}\!=\!\delta_{mN}\Delta_{0} \tilde{f}_{N}^{{\rm R}}$
with $\tilde{f}_{N}^{{\rm R}*}(\varepsilon,\theta)\!=\!\tilde{f}_{N}^{{\rm R}}(-\varepsilon,\theta)$.
It then follows from Eqs.\ (\ref{pair}) and  (\ref{Eilen-fR2}) that
$\delta\hspace{-0.3mm}f^{{\rm R}}_{mN}$ and 
$\delta\hspace{-0.3mm}\Delta_{N}$
may be written as 
\begin{subequations}
\begin{equation}
\delta\hspace{-0.3mm}\Delta_{N} = 
\Delta_{0}\delta_{N1}\delta\hspace{-0.3mm}\tilde{\Delta}_{1} \, ,
\label{Delta_N}
\end{equation}
\begin{equation}
\delta\hspace{-0.3mm}f^{{\rm R}}_{mN} = \Delta_{0}\left(
\delta_{m,N-1}\,\delta\hspace{-0.3mm}{\tilde f}^{{\rm R}}_{N+}
+\delta_{m,N+1}\,\delta\hspace{-0.3mm}{\tilde f}^{{\rm R}}_{N-}\right) \, .
\label{fR_mN}
\end{equation}
\end{subequations}
Equation (\ref{Eilen-fR2}) is thereby transformed into $(\mu\!=\!\pm)$
\begin{eqnarray}
\sum_{N'} {\cal M}_{NN'}\delta\hspace{-0.3mm}{\tilde f}^{{\rm R}}_{N'\mu} 
= ({\tilde f}^{{\rm R}}_{N+}+{\tilde f}^{{\rm R}}_{N-})
\,\delta E_{\mu}\,{\sin\theta}/{4\omega}
\nonumber \\
\hspace{3mm}+\delta_{N1}\delta_{\mu +}
\left(\delta\hspace{-0.3mm}\tilde{\Delta}_{1}
+{\langle \delta\hspace{-0.3mm}{\tilde f}^{{\rm R}}_{1+}\rangle}/{2\tau}\right) \, ,
\label{Eilen-fR3}
\end{eqnarray}
where $E_{\pm}\!\equiv\!E_{x}\!\pm\! iE_{y}$, 
${\tilde f}^{{\rm R}}_{N\pm}\!\equiv\! {\tilde f}^{{\rm R}}_{N}(\varepsilon_{\pm})$, 
and
\begin{subequations}
\label{df^R}
\begin{equation}
{\cal M}_{NN'}\equiv -i\tilde{\varepsilon}\,\delta_{NN'}
+\beta\sqrt{N\!+\!1}\,\delta_{N,N'-1}
-\beta\sqrt{N}\,\delta_{N,N'+1}\, ,
\label{Matrix}
\end{equation}
with $\tilde{\varepsilon}\!\equiv\!
\varepsilon\!+\!i/2\tau$ and $\beta\!\equiv\! \sqrt{B}\sin\theta/2\sqrt{2}$.
To solve Eq.\ (\ref{Eilen-fR3}), let us write the inverse of the matrix ${\cal M}$ as
\begin{eqnarray}
K^{N'}_{N}\equiv ({\cal M}^{-1})_{NN'} \, .
\label{KN'N}
\end{eqnarray}
Then $\delta\hspace{-0.3mm}{\tilde f}^{{\rm R}}_{N\mu}$ is obtained
formally as
\begin{equation}
\delta\hspace{-0.3mm}{\tilde f}^{{\rm R}}_{N\mu}
=K^{{\rm R}}_{N}{\delta E_{\mu}}/{\omega}+\delta_{\mu+}K^{1}_{N}
\left(\delta\hspace{-0.3mm}\tilde{\Delta}_{1}
+{\langle \delta\hspace{-0.3mm}{\tilde f}^{{\rm R}}_{1+}\rangle}/{2\tau}\right) \, ,
\label{df_N}
\end{equation}
where $K^{{\rm R}}_{N}$ is defined by
\begin{eqnarray}
K^{{\rm R}}_{N}\equiv \frac{\sin\theta}{4}
\sum_{N'}K_{N}^{N'} ({\tilde f}^{{\rm R}}_{N'+}+{\tilde f}^{{\rm R}}_{N'-})\, ,
\label{K^R}
\end{eqnarray}
with ${\tilde f}^{{\rm R}}_{N}$
already obtained as \cite{Kita03}
\begin{eqnarray}
f^{{\rm R}}_{N}=K^{0}_{N}/(1-\langle K^{0}_{0}\rangle/2\tau) \, .
\label{f^R_N}
\end{eqnarray}
Taking the angle average of Eq.\ (\ref{df_N}), we obtain
\begin{eqnarray}
\langle\delta\hspace{-0.3mm}{\tilde f}^{{\rm R}}_{1+}\rangle
=\frac{\langle K^{{\rm R}}_{1}\rangle}{1-\langle K_{1}^{1}\rangle/2\tau}
\frac{\delta E_{+}}{\omega}
+\frac{\langle K^{1}_{1}\rangle}{1-\langle K_{1}^{1}\rangle/2\tau}\delta\hspace{-0.3mm}
\tilde{\Delta}_{1}\, .
\label{df_1}
\end{eqnarray}
\end{subequations}
Equation (\ref{df^R}) determines
$\delta\hspace{-0.3mm}{\tilde f}^{{\rm R}}_{N\mu}$ efficiently
and fixes the retarded response $\delta\hspace{-0.3mm}f^{{\rm R}}$.

We next consider $\delta\hspace{-0.3mm}f^{\rm a}$
and substitute Eqs.\ (\ref{g^a(0)}), (\ref{fR-Expand}), and (\ref{fR_mN})
into Eq.\ (\ref{Eilen-fa}) with $g^{R}\!=\! -g^{A}\!=\! 1$. 
Expanding $\delta\hspace{-0.3mm}f^{\rm a}$ as Eq.\ (\ref{fR-Expand}),
we find that the coefficient $\delta\hspace{-0.3mm}f^{\rm a}_{mN}$ 
can also be written as Eq.\ (\ref{fR_mN}).
The corresponding $\delta\hspace{-0.3mm}{\tilde f}_{N\mu}^{\rm a}$
$(\mu\!=\!\pm)$ 
is obtained easily as
\begin{subequations}
\label{f^a0}
\begin{equation}
\delta\hspace{-0.3mm}{\tilde f}_{N\mu}^{\rm a}=K^{\rm a}_{N}
\frac{\phi_{+}-\phi_{-}}{\omega}\delta E_{\mu}\,,
\label{f^a}
\end{equation}
with
\begin{equation}
K^{\rm a}_{N}\equiv-\frac{\tau\sin\theta}{2(1-i\omega\tau)}
\left[{\tilde f}^{\rm R}_{N+}+(-1)^{N-1}{\tilde f}^{\rm R*}_{N-}\right] \, .
\label{K^a}
\end{equation}
\end{subequations}

Now we are ready to calculate the pair potential $\delta\hspace{-0.3mm}\Delta$.
Let us substitute the above results for $\delta\hspace{-0.3mm}{\tilde f}^{\rm R}$ and 
$\delta\hspace{-0.3mm}{\tilde f}^{\rm a}$ into Eq.\ (\ref{pair}).
Then a straightforward calculation yields the expression for 
$\delta\hspace{-0.3mm}{\tilde \Delta}_{1}$ defined by Eqs.\
(\ref{pairExpand}) and (\ref{Delta_N}) as
\begin{subequations}
\begin{equation}
\delta\hspace{-0.3mm}{\tilde \Delta}_{1}(\omega)=D(\omega)
\frac{\delta E_{+}}{\omega}\, ,
\label{Delta_1}
\end{equation}
with
\begin{equation}
D(\omega)=\frac{\displaystyle \frac{1}{2i}\!\int_{-\infty}^{\infty} \!\!
\left[\frac{\langle K^{\rm R}_{1}(\varepsilon,\omega)\rangle}
{1\!-\!\langle K^{1}_{1}(\varepsilon)\rangle/2\tau}
-\langle K^{\rm a}_{1}(\varepsilon,\omega)\rangle\right]\! 
\phi(\varepsilon_{-})\, d\varepsilon}
{\displaystyle \frac{1}{2}\!\int_{-\infty}^{\infty} \!\!\left\{\frac{1}{\sqrt{\varepsilon^{2}+1}}+
\frac{i\langle K^{\rm 1}_{1}(\varepsilon)\rangle}
{1\!-\!\langle K^{1}_{1}(\varepsilon)\rangle/2\tau}\,\phi(\varepsilon_{-})
\right\}d\varepsilon}\, .
\label{D}
\end{equation}
\end{subequations}
In deriving the formula, use has been made of the symmetries:
$K^{1*}_{N}(\varepsilon)\!=\!K^{1}_{N}(-\varepsilon)$, 
$K^{{\rm R}}_{N}(\varepsilon,\omega)
\!=\!K^{{\rm R}}_{N}(\varepsilon,-\omega)\!=\!K^{{\rm R}*}_{N}(-\varepsilon,\omega)$, 
and 
$K^{{\rm a}}_{N}(\varepsilon,\omega)\!=\!(-1)^{N-1}K^{{\rm a}}_{N}(-\varepsilon,\omega)
\!=\!K^{{\rm a}*}_{N}(-\varepsilon,-\omega)$.
I also substituted the result
$1/N(0)V_{0}\!=\!\frac{1}{2}\int_{-\varepsilon_{c}}^{\varepsilon_{c}} 
\frac{d\varepsilon}{\sqrt{\varepsilon^{2}+1}} $
at $T\!=\!H\!=\!0$ and put $\varepsilon_{c}\!\rightarrow\!\infty$,
noting $K^{1}_{1}(\varepsilon)\!\rightarrow\! i/\varepsilon$ as 
$\varepsilon\!\rightarrow\!\infty$, etc.
One can see easily that Eq.\ (\ref{D}) satisfies $D(-\omega)\!=\!D^{*}(\omega)$.

The second-order quantities $\delta g^{{\rm R,a}(2)}$
can be calculated similarly by expanding them as
\begin{equation}
\delta g^{\rm R,a(2)}(\varepsilon,{\bf k},\omega,{\bf r}) = \sum_{m}
\sum_{{\bf K}} \delta g_{m{\bf K}}^{{\rm R,a}(2)}(\varepsilon,\theta,\omega)\, 
{\rm e}^{i m\varphi+i{\bf K}\cdot{\bf r}} \, ,
\label{gRaExpand}
\end{equation}
where ${\bf K}$ is a reciprocal lattice vector of the magnetic Brillouin zone \cite{Kita98}.
Since $\delta{\bf E}$ is spatially uniform, we only need the ${\bf K}\!=\!{\bf 0}$ component
within the linear-response regime.
It also follows from Eq.\ (\ref{current}) that $m\!\neq \!\pm 1$ are irrelevant for the
current density.
Let us substitute the expansions for $f^{\rm R}$ and $\delta f^{\rm R}$
into Eq.\ (\ref{Eilen-gR}) with $g^{R}\!=\!1$,
multiplies it by ${\rm e}^{\mp i\varphi}/2\pi V$,
and perform integrations over $({\bf r},\varphi)$.
We thereby obtain
$g_{\pm}^{{\rm R}(2)}\!\equiv\!g_{m=\pm 1,{\bf K}={\bf 0}}^{{\rm R}(2)}$ as
\begin{subequations}
\label{g^R}
\begin{equation}
\delta g^{\rm R(2)}_{\pm}(\varepsilon,\omega)=\Delta_{0}^{2}\,
\Gamma^{\rm R}_{\pm}(\varepsilon,\omega)\,\frac{\delta E_{\mp}}{\omega} \, ,
\label{g^R1}
\end{equation}
with ($\mu\!=\!\pm$)
\begin{eqnarray}
&&\hspace{-10mm}\Gamma^{\rm R}_{\mu}
\equiv -\frac{1}{2}\sum_{N}(-1)^{N}\biggl[({\tilde f}^{{\rm R}}_{N+}+{\tilde f}^{{\rm R}}_{N-})
K^{\rm R}_{N}
\nonumber \\
&&\hspace{-2mm}+({\tilde f}^{{\rm R}}_{N+}\delta_{\mu +}+
{\tilde f}^{{\rm R}}_{N-}\delta_{\mu -})K^{1}_{N}\,
\frac{\langle K^{\rm R}_{1}\rangle/2\tau+D}{1-\langle K^{1}_{1}\rangle/2\tau}\,\biggr] .
\label{g^R2}
\end{eqnarray}
\end{subequations}
The quantity $g_{\pm}^{{\rm a}(2)}\!\equiv\!g_{m=\pm 1,{\bf K}={\bf 0}}^{{\rm a}(2)}$
can be obtained similarly from Eq.\ (\ref{Eilen-ga}) as
\begin{subequations}
\label{g^a}
\begin{equation}
\delta g^{\rm a(2)}_{\pm}(\varepsilon,\omega)=\Delta_{0}^{2}\,
\Gamma^{\rm a}_{\pm}(\varepsilon,\omega)
\frac{\phi(\varepsilon_{+})-\phi(\varepsilon_{-})}{\omega}\delta E_{\mp} \, ,
\label{g^a1}
\end{equation}
with
\begin{eqnarray}
&&\hspace{-5mm}\Gamma^{\rm a}_{\mu}
\equiv \frac{\tau\sin\theta}{4(1-i\omega\tau)}\sum_{N}(-1)^{N}
\bigl[ ({\tilde f}^{{\rm R}}_{N+})^{2} +
       ({\tilde f}^{{\rm R}*}_{N-})^{2}\bigr]
\nonumber \\
&&-\frac{\tau\sin\theta}{4(1-i\omega\tau)^{2}}\sum_{N}(-1)^{N}
\bigl< ({\tilde f}^{{\rm R}}_{N+})^{2}+ ({\tilde f}^{{\rm R}*}_{N-})^{2}\bigr>
\nonumber \\
&&-\frac{\bigl< {\tilde f}^{{\rm R}}_{0+}- {\tilde f}^{{\rm R}*}_{0-}\bigr>
K^{\rm a}_{0}
+\bigl({\tilde f}^{{\rm R}}_{1+}\delta_{\mu +}
+{\tilde f}^{{\rm R}*}_{1-}\delta_{\mu -}\bigr)\bigl< K^{\rm a}_{1}\bigr>}
{2(1-i\omega\tau)}
\nonumber \\
&&
-\frac{ \tau\bigl({\tilde f}^{{\rm R}}_{1+}\delta_{\mu +}
+{\tilde f}^{{\rm R}*}_{1-}\delta_{\mu -}\bigr)}{(1-i\omega\tau)}
D \, .
\label{g^a2}
\end{eqnarray}
\end{subequations}

Substituting Eqs.\ (\ref{gRaExpand})-(\ref{g^a}) into Eq.\ (\ref{current})
and noting $\Gamma^{{\rm R}}_{\mu}(\varepsilon,\omega)
\!=\!\Gamma^{{\rm R}*}_{\mu}(-\varepsilon,-\omega)$ and 
$\Gamma^{{\rm a}}_{\mu}(\varepsilon,\omega)
\!=\!\Gamma^{{\rm a}}_{-\mu}(-\varepsilon,\omega)
\!=\!\Gamma^{{\rm a}*}_{\mu}(-\varepsilon,-\omega)$,
we finally obtain $\delta j^{(2)}_{\pm}\!\equiv\! j^{(2)}_{x}\!\pm i j^{(2)}_{y}$ as
\begin{subequations}
\begin{equation}
\delta j^{(2)}_{\pm}(\omega)=\sigma_{f}^{(2)}(\omega)\, \delta\hspace{-0.3mm}E_{\pm} \, ,
\label{current2}
\end{equation}
with
\begin{eqnarray}
&&\hspace{-10mm}\sigma_{f}^{(2)}(\omega)
\equiv-\Delta_{0}^{2}\frac{eN(0)v_{\rm F}}{2\omega}\!\!\!
\int_{-\infty}^{\infty}d\varepsilon\, \phi(\varepsilon_{-})
\nonumber \\ 
&& \hspace{6mm}\times\sum_{\mu}
\bigl<\!\bigl[\Gamma^{\rm R}_{\mu}(\varepsilon,\omega)
-\Gamma^{\rm a}_{\mu}(\varepsilon,\omega)\bigr]\sin\theta\bigr>
\, .
\label{dSigma}
\end{eqnarray}
\end{subequations}
Thus $\sigma_{f}^{(2)}$ is isotropic,
as expected for any second-order corrections,
and satisfies $\sigma_{f}^{(2)}(-\omega)\!=\!\sigma_{f}^{(2)*}(\omega)$.

Equation (\ref{dSigma}) forms the main result of the paper,
which is not only exact but also convenient for numerical calculations.
The key quantities are $K^{N'}_{N}$'s defined by Eq.\ (\ref{KN'N}) with Eq.\ (\ref{Matrix}),
as may be seen from Eqs.\ (\ref{K^R}), (\ref{f^R_N}), (\ref{K^a}), (\ref{D}),
(\ref{g^R2}), and (\ref{g^a2}), satisfying $K^{N'}_{N}\!=\!(-1)^{N+N'}K^{N}_{N'}$.
An efficient algorithm to calculate them is obtained as follows.
Let us define ${\cal D}_{N}$ ($\tilde{{\cal D}}_{N}$)
for $N\!=\!0,1,2,\cdots$ as the determinant of the submatrix 
obtained by removing (retaining) the first $N$ rows and columns 
of the tridiagonal matrix ${\cal M}$, Eq.\ (\ref{Matrix}).
They satisfy ${\cal D}_{N-1}\!=\!-i\tilde{\varepsilon}{\cal D}_{N}\!+\! \beta^{2}N {\cal D}_{N+1}$
and ${\tilde {\cal D}}_{N+1}\!=\!-i\tilde{\varepsilon}{\tilde {\cal D}}_{N}
\!+\! \beta^{2}N{\tilde {\cal {\cal D}}}_{N-1}$
with ${\tilde {\cal D}}_{1}\!=\!-i\tilde{\varepsilon}$ and ${\tilde {\cal D}}_{0}\!=\!1$,
as can be shown with standard matrix manipulations.
Then $K^{N'}_{N}$ for $N'\!\leq\! N$ is obtained,
by also using standard techniques to invert a matrix,
as $K^{N'}_{N}\!=\!\beta^{N-N'}(N!/N'!)^{1/2}{\cal D}_{N+1}{\tilde {\cal D}}_{N'}/{\cal D}_{0}$.

This algorithm can be put into a more convenient form in terms of 
${\cal R}_{N}\!\equiv\! {\cal D}_{N+1}/{\cal D}_{N}$ and ${\tilde {\cal R}}_{N}
\!\equiv\! {\tilde {\cal D}}_{N-1}/{\tilde {\cal D}}_{N}$ 
as follows.
Let us calculate ${\cal R}_{N}$ and ${\tilde {\cal R}}_{N}$ by
\begin{subequations}
\label{RK}
\begin{equation}
\left\{
\begin{array}{ll}
{\cal R}_{N-1}=(-i\tilde{\varepsilon}+\beta^{2}N {\cal R}_{N})^{-1} ,
&\hspace{3mm} {\cal R}_{N_{\rm{cut}}}=i/\tilde{\varepsilon} \, , \\
{\tilde {\cal R}}_{N+1}=(-i\tilde{\varepsilon}+\beta^{2}N {\tilde {\cal R}}_{N})^{-1},
&\hspace{3mm} {\tilde {\cal R}}_{1}=i/\tilde{\varepsilon} \, ,
\end{array}
\right.
\label{R_N}
\end{equation}
for an appropriately chosen large $N_{\rm{cut}}$.
Then $K^{N'}_{N}$ for $N'\!\leq\! N$ is obtained by
\begin{equation}
\left\{
\begin{array}{l}
K^{0}_{0}={\cal R}_{0}, 
\hspace{5mm}K^{N'}_{N'}=({\cal R}_{N'}/{\tilde {\cal R}}_{N'})K^{N'-1}_{N'-1}\, ,\\
K^{N'}_{N+1}=\beta \sqrt{N+1}{\cal R}_{N+1} K^{N'}_{N}\, . 
\end{array}
\right.
\label{KN'N2}
\end{equation}
\end{subequations}
One can check the convergence by increasing $N_{\rm cut}$.
It turns out that
$N_{\rm cut}\!=\! 2$ is sufficient both near $T_{c}$
and in the dirty limit where an analytic calculation is also possible.
One can thereby reproduce the formula by Thompson 
\cite{Thompson70} and Takayama and Ebisawa \cite{TE70}
which satisfies ${\rm Im}\,\sigma_{f}^{(2)}(\omega\!\rightarrow \!0)\!=\! 0$.
In contrast, $N_{\rm cut}\!\agt\! 1000$
is required in the clean limit at low temperatures.

\begin{figure}[t]
\includegraphics[width=0.8\linewidth]{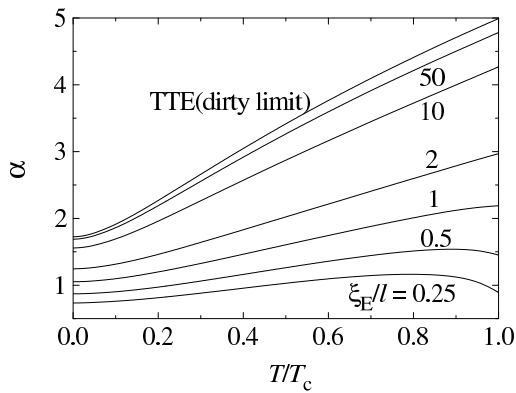}%
\caption{The slope $\alpha$ of the flux-flow resistivity at $H_{c2}$
as a function of the reduced temperature $T/T_{c}$
for several values of $\xi_{\rm E}/l \!\equiv\! 1/2\pi T_{c}\tau$
with $\kappa_{\rm GL}\!=\!50$.
The curve TTE denotes the prediction of the dirty-limit theory \cite{Thompson70,TE70}.
}
\label{fig:1}
\end{figure}

Using Eq.\ (\ref{dSigma}) and taking the limit $\omega\!\rightarrow\!0$, 
I have calculated the initial slope of the flux-flow resistivity:
\begin{equation}
\alpha\equiv \frac{H}{\rho_{n}}\left.\frac{\partial \rho_{f}}{\partial H}\right|_{H=H_{c2}} \, ,
\label{alpha}
\end{equation}
for various impurity concentrations and various values of the Ginzburg-Landau parameter 
$\kappa_{\rm GL}$.
The main $H$ dependence in Eq.\ (\ref{dSigma}) lies in 
$\Delta_{0}^{2}\!\propto\!H_{c2}\!-\!H$,
which can be obtained accurately following Ref.\ \onlinecite{Kita03}.

Figure 1 displays the slope $\alpha$
for several values of $\xi_{\rm E}/l \!\equiv\! 1/2\pi T_{c}\tau$
with $\kappa_{\rm GL}\!=\!50$.
The curve TTE denotes the prediction of the dirty-limit theory
by Thompson \cite{Thompson70} and Takayama and Ebisawa \cite{TE70}.
Marked mean-free-path dependence is clearly seen.
In fact, 
the slope at $T\!=\!T_{c}$ ($T\!=\!0$) decreases from $4.99$ ($1.72$) in the dirty limit to
$0.89$ ($0.73$) at $\xi_{\rm E}/l \!=\!4.0$.
Thus, nonlocal effects in clean systems tend to increase
the resistivity substantially over the prediction of the dirty-limit theory.
This result near $H_{c2}$ also suggests that $\rho_{f}(H)$ at low temperatures
may change from convex downward to upward as $l$ increases
and may not be fit quantitatively by
the Bardeen-Stephen theory $\rho_{f}\!\propto\! H/H_{c2}$ \cite{BS65}.
The slope has also been found to become steeper 
as $\kappa_{\rm GL}$ approaches $1/\sqrt{2}$,
due to the increase of the coefficient of $\Delta_{0}^{2}\!\propto\!H_{c2}\!-\!H$, 
reaching $\alpha\!=\!9.57$ ($2.59$) at $T\!=\!T_{c}$ ($T\!=\!0$) for $\kappa_{\rm GL}\!=\! 1$
and $\xi_{\rm E}/l \!=\!50$.
This fact indicates the necessity of correctly identifying the material parameters,
such as $\kappa_{\rm GL}$,
$\xi$, and $l$, for any detailed comparisons
between the theory and experiments on the flux-flow resistivity.
Besides, a careful experiment will be required, 
especially near $T_{c}$,
to determine the slope $\alpha$ 
which may change appreciably near $H_{c2}$ \cite{PKT73}.

In summary, this paper has developed a reliable and efficient method to calculate the
flux-flow resistivity near $H_{c2}$ over all impurity concentrations
and clarified large dependence of $\rho_{f}$ on both $l$ and $\kappa_{\rm GL}$.

This research is supported by Grant-in-Aid for Scientific Research 
from the Ministry of Education, Culture, Sports, Science, and Technology
of Japan.


\end{document}